\documentclass[11pt]{article}
\usepackage{amsmath,amsfonts,sint,epsfig,cite,graphics}
\usepackage{mathrsfs}
\usepackage{macros_static}
\usepackage{wrapfig}
\usepackage{color}

\usepackage[english]{babel}

\usepackage{psfrag}
\usepackage{amsmath}
\usepackage{dsfont}
\usepackage{subfigure}
\usepackage{axodraw}
\usepackage{feynmp}
\usepackage{amsfonts}

\newcommand{\bdm}{\begin{displaymath}}
\newcommand{\edm}{\end{displaymath}}
\renewcommand{\be}{\begin{equation}}
\renewcommand{\ee}{\end{equation}}
\renewcommand{\bi}{\begin{itemize}}
\renewcommand{\ei}{\end{itemize}}

\newcommand{\Str}{{\rm Str}}
\newcommand{\str}{{\rm Str}}
\newcommand{\mpis}{{m_\pi^2}}
\newcommand{\metas}{{m_\eta^2}}
\newcommand{\mKs}{{m_K^2}}

\newcommand{\Abar}{{\bar{A}}}
\newcommand{\Bbar}{{\bar{B}}}

\include{include}

\begin{document}

\renewcommand{\vec}[1]{{\bf #1}}
\newcommand{\PH}{P_{\rm H}}
\newcommand{\PL}{P_{\rm L}}
\newcommand{\unity}{{I}}

\def\slash#1{\mkern-1.5mu\raise0.6pt\hbox{$\not$}\mkern1.2mu #1\mkern 0.7mu}

\begin{titlepage}

\begin{flushright}
\vskip .4cm
CERN-PH-TH/2011-254\\
\end{flushright}

\vskip 2.25cm
\begin{center}
{\Large\bf 
Revisiting the pion's scalar form factor in chiral perturbation theory}
\end{center}
\vskip 1.1cm
\begin{center}
{
Andreas J\"uttner\\
}
{
CERN, Physics Department, TH Unit, CH-1211 Geneva 23, Switzerland
}
\end{center}
\vskip 1.3cm
{{\bf Abstract:} 
The quark-connected and the quark-disconnected Wick contractions contributing
to the pion's scalar form factor are computed in the two and in the three flavour
chiral effective theory at next-to-leading order. 
While the quark-disconnected contribution to the form factor itself turns out to be
power-counting suppressed its contribution to the 
scalar radius is of the same order of magnitude 
as the one of the quark-connected contribution.
This result underlines that neglecting quark-disconnected contributions in 
simulations of lattice QCD can cause significant systematic effects. The technique
used to derive these predictions can be applied to a large class of observables 
relevant for QCD-phenomenology.

}

\end{titlepage}
\newpage
\tableofcontents
\section{Introduction}
At first sight the scalar form factor of the pion appears to be
 a purely academic
quantity since despite being a well-defined object in QCD, nature does not
seem to provide a low-energy scalar probe.
The experimental information on the scalar $I=0$ $\pi\pi$ phase
shift however 
serves as an input parameter in the dispersive representation of the
form factor. Indirectly a comparison with
 experiment is therefore in principle possible.

Lattice QCD is the tool of choice for determining low energy
properties of hadrons~\cite{Colangelo:2010et}. 
However, while computing the scalar form factor  
is in principle a straight forward procedure it turns out 
to be numerically much more complicated than the computation of the closely related
pion vector form factor in the iso-spin limit
\cite{Capitani:2005ce,Boyle:2008yd,Frezzotti:2008dr,Aoki:2009qn}. 
The reason lies in the contribution of 
quark-disconnected Wick contractions to its defining correlation functions
in the scalar case. 
There is no conceptual difficulty in treating these \cite{Aoki:2009qn}
but the computational effort is immense
compared to quark-connected contributions. The disconnected part is 
therefore often neglected, not always providing supporting arguments that the
systematic effect introduced in this way is under control.

This paper provides the expressions for the contributions of  the
 quark connected and the quark disconnected 
Wick contractions to the scalar form factor of the pion 
in chiral perturbation theory at next-to-leading order 
(NLO)~\cite{Gasser:1983yg,Gasser:1984gg,Gasser:1984ux}. 
As one might have expected and as first numerical studies indicate \cite{Aoki:2009qn},
the relative contribution of the quark-disconnected part to the form factor in 
the space-like region turns out
to be small for small values of the momentum transfer. 
However, its contribution to the form factor above threshold and to the scalar
 radius which is defined as the form factor slope
at vanishing momentum transfer turn out to be of about the same magnitude as the one
of the quark-connected contribution. 

While underlining the potential shortcomings of neglecting quark-disconnected contributions
in lattice computations of QCD observables the results derived in this work 
allow to predict the
disconnected contribution at NLO in the effective theory. 
Moreover,
the arguments presented here justify the use of (partially) twisted 
boundary conditions in a lattice computation 
\cite{Sachrajda:2004mi,Bedaque:2004ax,Bedaque:2004kc,deDivitiis:2004kq,
	Tiburzi:2005hg,Flynn:2005in}
allowing to compute the
form factor on the finite lattice volume for arbitrary values of the 
momentum transfer.

The technique used to derive these predictions has recently been introduced
in \cite{DellaMorte:2010aq} where it was applied to the two-point correlator
of vector currents in QCD. After a brief summary of the idea the 
important steps in the computation for the scalar form factor within this frame work
are presented for the theory with both 
$N_f=2$ and $N_f=2+1$ flavours, followed by an illustration
and discussion of the results.

\section{Quark disconnected diagrams in chiral effective theory}\label{sec:general_argument}
\begin{figure}
    \begin{center}
    \begin{picture}(80,40)(-40,-11)
	\GOval(0,0)(10,20)(0){1}
	\GCirc(0,10){2}{.3}
	\GCirc(-20,0){2}{.3}
	\GCirc(20,0){2}{.3}
	\Text(-30,0)[c]{$\mathcal{O}$}
	\Text(+30,0)[c]{$\mathcal{O}$}
	\Text(0,20)[c]{$S$}
    \end{picture}\hspace{1.5cm}
    \begin{picture}(80,40)(-40,-11)
	\GCirc(0,19){7}{1}
	\GCirc(0,26){2}{.3}
	\GOval(0,0)(10,20)(0){1}
	\GCirc(-20,0){2}{.3}
	\GCirc(20,0){2}{.3}
	\Text(-30,0)[c]{$\mathcal{O}$}
	\Text(+30,0)[c]{$\mathcal{O}$}
	\Text(0,33)[c]{$S$}
    \end{picture}
    \end{center}
 \caption{Wick contractions contributing to the scalar form factor.}
	\label{fig:Wicks}
\end{figure}
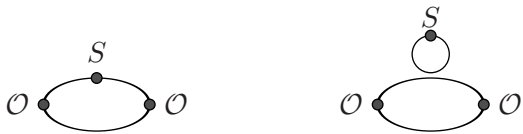

The technique will be briefly discussed for the example of the 2-flavour theory. The 
generalisation to the 2+1-flavour theory is straight forward and will become clear in the
following sections.
The scalar form factor of the pion is defined through
\begin{equation}\label{eq:SU2ME}
\langle \pi^i(p^\prime)|\bar u u+\bar d d|\pi^j(p)\rangle=
	\delta^{ij}F_{S,2}(t)\,.
\end{equation}
where $t=(p^\prime-p)^2$ is the squared momentum transfer and where the sub-script 2
indicates $N_f=2$ flavours.
The matrix element on the 
l.h.s. is the ground state contribution to the
Fourier transform of the QCD quark three point  function 
\begin{equation}
 \left< \mathcal{O}^i(z)S(y){\mathcal{O}^i}^\dagger(x)\right>\,,
\end{equation}
constructed of the interpolating operators
$O^i(x)=\bar \psi_2 \tau^i\gamma_5 \psi_2(x)$ and 
$S(x)=\bar \psi_2(x)          \psi_2(x)$.  The $\psi^T_2=(u,d)$ are  
$SU(2)$ flavour vectors of $u$- and $d$-quarks and 
the matrices $\tau_i=\sigma_i/2$ are proportional to the Pauli matrices.
On the level of the quark fields there are two classes of Wick-contractions
as illustrated in figure~\ref{fig:Wicks}.

As detailed in \cite{DellaMorte:2010aq},
by  introducing an additional valence quark, $v$, which is degenerate 
with the light 
dynamical flavours, each Wick contraction can be rewritten in terms
of a single fermionic correlation function defined in an un-physical 
theory. The 
physical result is recovered by summing over the correlation functions 
in the un-physical, partially quenched, theory 
\cite{Bernard:1992mk,Bernard:1993sv,Sharpe:2000bc}. In the case we consider 
here, e.g., 
\begin{equation}\label{eq:decomp_quarks}
\langle \bar u \gamma_5 d\, \bar d d\, \bar d \gamma_5 u\rangle=
\langle \bar u \gamma_5 v\, \bar v d\, \bar d \gamma_5 u\rangle+
\langle \bar u \gamma_5 d\, \bar v v\, \bar d \gamma_5 u\rangle\,.
\end{equation}
An expression for the l.h.s. in the effective theory has been 
found many years ago at NLO~\cite{Gasser:1983yg,Gasser:1984ux} and 
at NNLO in~\cite{Bijnens:1998fm,Bijnens:2003xg} for both the $N_f=2$ and
$N_f=2+1$ theory.
This paper provides expressions for the ground state matrix elements
of the terms on the r.h.s. of eq.~(\ref{eq:decomp_quarks}) 
in partially quenched chiral perturbation theory (PQ$\chi$PT) 
\cite{Bernard:1992mk,Bernard:1993sv,Sharpe:2000bc}, which is an extension
of chiral perturbation theory \cite{Gasser:1984gg,Gasser:1983yg} that
provides 
an asymptotic low energy description of partially quenched QCD (PQQCD).
Following~\cite{DellaMorte:2010aq}, the
flavour group is promoted to the graded $SU(3|1)$ 
with the flavour vector $\psi_{3|1}^T=(u,d,v,g)$ extended
by an additional valence and ghost quark. Both are assumed to be 
mass degenerate to the up- and down quarks.  
The generators of the thus enlarged flavour group are $T^a$, 
$a=1,\dots,15$.
We use the conventions also employed in \cite{Giusti:2008vb,DellaMorte:2010aq},
\be
T^a=(T^a)^\dagger\;, \quad \quad \Str\left\{T^a\right\}=0\;, \quad
\Str\left\{T^aT^b\right\}={{1}\over{2}}g^{ab}\;, \quad a=1,\dots,15\,,
\ee
with the \textit{super-trace}
$\Str\left\{A\right\}=A_{11}+A_{22}+A_{33}-A_{44}$.
The matrix $g^{ab}$ reads
\begin{equation}
  g=\begin{pmatrix}1\cr
               &\ddots\cr
               &      &1\cr
               &      & &-\sigma_2\cr
               &      & &       &-\sigma_2\cr
               &      & &       &      &-\sigma_2\cr
               &      & &       &      &       &-1\cr\end{pmatrix}
  \begin{matrix}\left.\vphantom{\begin{matrix}1\cr
                                   &\ddots\cr
                                   &      &1\cr \end{matrix}}\right\}&
                                   \kern-1.5ex1-8\hfill\cr
          \left.\vphantom{\begin{matrix}-\tau^2\cr
                                         &\ddots\cr
                                         &      &-\tau^2\cr\end{matrix}}\right\}&
                                   \kern-1.5ex9-14\hfill\cr
          \left.\vphantom{\begin{matrix} 1\cr \end{matrix} }\right\}&
                                   \kern-1.5ex15\hfill\cr
         \end{matrix}
\end{equation}
where $\sigma_2$ is the second Pauli matrix. $T^1,\dots, T^8$ are the 
generators of the $SU(3)$ subgroup that acts on the sea and valence 
components, $T^9,\dots, T^{14}$ mix the quark with the ghost components,
$T^{15}$ is the diagonal matrix ${\rm diag}(1,1,1,3)/(2\sqrt{3})$.
The matrix $T^0$ with $\Str\{T^0\}=1/\sqrt{2}$
is proportional to the unit matrix.

In this theory the matrix element in eq.~(\ref{eq:SU2ME}) can be written as
\begin{equation}\label{eq:decomposition}
\begin{array}{rcl}
 \Big< \pi^1\Big|S^0+\frac{S^8-S^{15}}{\sqrt{3}}\Big|{\pi^1}\Big>&=&
 \Big< \pi^1\Big|2S^4\Big|{\pi^1}\Big>+
 \Big< \pi^1\Big|S^0-\frac{2S^8+S^{15}}{\sqrt{3}}\Big|{\pi^6}\Big>\,,
\end{array}
\end{equation}
where the scalar source  now is of 
the form $S^i=\bar\psi_{3|1} T^i\psi_{3|1}$.

All previous steps can also be applied to the 2+1-flavour theory. There, the $SU(3)$ flavour
symmetry is promoted to a graded $SU(4|1)$ and the flavour content of the theory is
represented by the flavour vector $\psi^T_{4|1}=(u,d,s,v,g)$ with $s$ being the strange
quark field. The graded group $SU(4|1)$ has 24 generators and the super-trace as well as
 the metric $g^{ab}$ and $T^0$ 
introduced above for $SU(3|1)$ need to be modified correspondingly.

In the next section the chiral effective theory corresponding to 
the $SU(3|1)$ and $SU(4|1)$ flavour symmetry groups is set up.  

\section{PQ$\chi$PT for the scalar pion form factor}
The degrees of freedom of the partially quenched theory are 
parameterised in terms of 
 $U=\exp\left(2i\frac {\phi^a T^a} F\right)$, where 
the $\phi^a$ are the Goldstone-boson/fermion fields ($a=1,\dots,15$
for $SU(3|1)$ and $a=1,\dots,24$ for $SU(4|1)$).
All vertices relevant for the computation at NLO of the scalar form factor
(illustrated in figure \ref{fig:feyn_a}-\ref{fig:feyn_c} and 
explicitly given in appendix \ref{app:vertices}) 
can be derived from the  
leading order chiral Lagrangian 
\cite{Gasser:1984gg,Gasser:1983yg,Bernard:1992mk,Bernard:1993sv,Sharpe:2000bc}
\begin{equation}\label{eq:L2}
\mathcal{L}^{(2)}=\frac{F^2}{4}\Str\left\{
	\partial_\mu U \partial^\mu U^\dagger\right\}
	+\frac {F^2}4  \Str\left\{\chi U^\dagger + U\chi^\dagger \right\}\,,
\end{equation}
where $\chi=2 B (s+M)$. The mass matrix has the form 
\begin{equation}
\begin{array}{ccc}
SU(3|1)&&SU(4|1)\\
M={\rm diag}\left(m_q,m_q,m_q,m_q\right)\,,&\qquad&
M={\rm diag}\left(m_q,m_q,m_s,m_q,m_q\right)\,.
\end{array}
\end{equation} 
and we define the external scalar source as $s=2T^a s^a$.
The relevant counter terms illustrated in figure \ref{fig:feyn_d}  can be
derived from the Lagrangian ,
\begin{equation}\label{eq:L4}
\begin{array}{rcl}
\mathcal{L}^{(4)}&=&
	L_4\Str\Big\{\partial_\mu U (\partial^\mu U)^\dagger\Big\}
		\Str\Big\{\chi U^\dagger + U \chi^\dagger\Big\}	
	+L_5\Str\Big\{\big(\partial_\mu U (\partial^\mu U)^\dagger\big)
		\big(\chi U^\dagger+U\chi^\dagger\big)\Big\}\\[2mm]
	&&+L_6\Str\Big\{\chi U^\dagger + U\chi^\dagger\Big\}^2
	+L_8\Str\Big\{U\chi^\dagger U\chi^\dagger + \chi U^\dagger \chi U^\dagger\Big\}^2\,.
\end{array}
\end{equation}
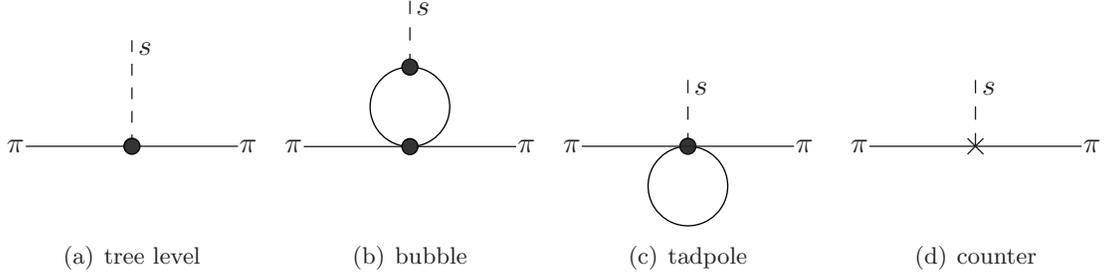
\begin{figure}
 \begin{center}
 \subfigure[tree level]{
    \begin{picture}(80,40)(-40,-30)
    \DashLine(0,0)(0,40){5}
    \Line(-40,0)(40,0)
    \GCirc(0,0){3}{.2}
    \Text(-44,0)[c]{$\pi$}
    \Text(+44,0)[c]{$\pi$}
    \Text(5,+37)[c]{$s$}
    \end{picture}
    \label{fig:feyn_a}
  }\hspace{5mm}
  \subfigure[bubble]{
    \begin{picture}(80,40)(-40,-45)
    \GCirc(0,0){15}{10}
    \DashLine(0,15)(0,40){5}
    \GCirc(0,15){3}{.2}
    \Line(-40,-15)(40,-15)
    \GCirc(0,-15){3}{.2}
    \Text(-44,-15)[c]{$\pi$}
    \Text(+44,-15)[c]{$\pi$}
    \Text(5,+37)[c]{$s$}
    \end{picture}
    \label{fig:feyn_b}
   }\hspace{5mm}
  \subfigure[tadpole]{
    \begin{picture}(80,40)(-40,-15)
    \GCirc(0,0){15}{10}
    \DashLine(0,15)(0,40){5}
    \GCirc(0,15){3}{.2}
    \Line(-40,+15)(40,+15)
    \Text(-44,+15)[c]{$\pi$}
    \Text(+44,+15)[c]{$\pi$}
    \Text(5,+37)[c]{$s$}
    \end{picture}
    \label{fig:feyn_c}
   } \hspace{5mm} 
  \subfigure[counter]{
    \begin{picture}(80,40)(-40,-15)
    \DashLine(0,15)(0,40){5}
    \Line(-40,+15)(40,+15)
    \Line(-3,12)(+3,+18)
    \Line(-3,18)(+3,+12)
    \Text(-44,+15)[c]{$\pi$}
    \Text(+44,+15)[c]{$\pi$}
    \Text(5,+37)[c]{$s$}
    \end{picture}
    \label{fig:feyn_d}
   }
  \end{center}
\caption{Contributing diagrams in the effective theory. Solid lines
	represent \textit{pions} built of fermionic or bosonic quarks
	and the dashed line represents the external scalar source.
	}\label{fig:diagrams}
\end{figure}
The partially quenched propagators are constructed in analogy to the procedure
presented in detail in~\cite{Giusti:2008vb,DellaMorte:2010aq}. Note however
that in the current paper we use Minkowski-signature.
\section{Results for the form factor}
\subsection{Expressions in the effective theory at NLO}
For the 2-flavour theory the expressions for the form factor are
\begin{equation}\label{eq:results}
\begin{array}{l@{\hspace{1mm}}c@{\hspace{1mm}}l@{\hspace{0mm}}l@{\hspace{0mm}}l@{\hspace{0mm}}l@{\hspace{0mm}}l@{\hspace{0mm}}l@{\hspace{0mm}}l@{\hspace{0mm}}l@{\hspace{0mm}}}
F^{\rm F}_{S,2}(t)&=&
  2 B \Big\{1 +\frac 1{F^2}\Big(-\frac 1{2} \Abar(\mpis)+ \Lambda^{\rm F}_2&  
      +\frac 12 (2 &t - \;\,\mpis) \Bbar(\mpis,t) 
	&\Big)\Big\}\,,\\[3mm]
F^{\rm C}_{S,2}(t) &=& 
  2 B \Big\{1 +\frac 1{F^2}\Big(-\frac 1{2} \Abar(\mpis)+ \Lambda^{\rm C}_2& 
     +\frac 12 (&t - 2\mpis) \Bbar(\mpis,t) 
	 &\Big)\Big\}\,,\\[3mm]
F^{\rm D}_{S,2}(t)&=& 
  2 B \Big\{0 +\frac 1{F^2}\Big( \hspace{23mm} \Lambda^{\rm D}_2&
     +\frac 12 (&t + \;\,\mpis) \Bbar(\mpis,t) 
	 &\Big)\Big\}\,,
\end{array}
\end{equation}
where,
\begin{equation}\label{eq:results_uds}
\hspace{-1.6mm}\begin{array}{l@{\hspace{1mm}}c@{\hspace{1mm}}l@{\hspace{0mm}}l@{\hspace{0mm}}l@{\hspace{0mm}}l@{\hspace{0mm}}l@{\hspace{0mm}}l@{\hspace{0mm}}l@{\hspace{0mm}}}
 \Lambda_2^{{\rm F}}&=&
	 4\Big\{
	\mpis(-8 \tilde L_4^r-4 \tilde L_5^r&+16 \tilde L_6^r &+8\tilde L_8^r)
	&+t    (2  \tilde L_4^r+ \tilde L_5^r)
	&\Big\}\,,\\[3mm]
\Lambda_2^{{\rm C}}&=&
	 4\Big\{
	\mpis(-4 \tilde L_4^r-4 \tilde L_5^r&+8 \tilde L_6^r &+8\tilde L_8^r)
	&+t     \hspace{12.5mm}\tilde L_5^r
	&\Big\}\,,\\[3mm]
\Lambda_2^{{\rm D}}&=&
	 4\Big\{
	\mpis (-4 \tilde L_4^r &+ 8 \tilde L_6^r)&
	&+ t\;\,2      \tilde L_4^r
	&\Big\}\,.
\end{array}
\end{equation}
and where  {$t=(p{\,^\prime}-p)^2$} is the squared 
momentum transfer between the two pions.  
The kinematical functions $\bar A(m^2)$ and $\bar B(m^2,t)$ are defined 
in appendix~\ref{app:kinfunctions}. 
The superscripts F, C and D indicate the Full form factor and the Connected and
Disconnected contributions, respectively and 
the superscript $r$ indicates the subtraction of divergences in the
$\overline{MS}$-scheme at subtraction scale $\mu=0.77$GeV. A tilde
was added to the low-energy constants in order to distinguish them from 
the 3-flavour ones.
With
\begin{equation}
\mathcal{A}=\frac 16 \Abar(\metas)-\frac 12 \Abar(\mpis)\,,
\end{equation}
the results for the $N_f=2+1$ theory are\\
\begin{minipage}{\linewidth}
\begin{equation}\label{eq:results_ud}
\hspace{-.10cm}\begin{array}{r@{\hspace{1mm}}c@{\hspace{1mm}}l@{\hspace{0mm}}l@{\hspace{0mm}}l@{\hspace{0mm}}l@{\hspace{0mm}}l@{\hspace{0mm}}l@{\hspace{0mm}}l@{\hspace{0mm}}}
 F_{S,3}^{{\rm F}}(t)&=&
	2B\Big\{
	1+\frac{1}{F^2}\Big(\mathcal{A}&+\Lambda_3^{{\rm F}}\\[2mm]
	&&&\hspace*{-3cm} +\frac{\mpis}{18}\Bbar(\metas,t)
	&\hspace{-15mm}+\frac 12(2t\,\,- {\mpis})\Bbar(\mpis,t)
	&+\frac t4 \Bbar(\mKs,t)
	&
	&\Big)\Big\}\,,\\[3mm]
 F_{S,3}^{{\rm C}}(t)&=&
	2B\Big\{
	1+\frac 1{F^2}\Big(\mathcal{A}&
	+\Lambda_3^{{\rm C}}\\[2mm]
	&&&\hspace*{-3cm}\mbox{}
	&\hspace{-15mm} +\frac 12 (\,\,t- 2\mpis )\Bbar(\mpis,t)
	&+\frac t4 \Bbar(\mKs ,t)
	&+\frac {\mpis}3 \Bbar(\metas,\mpis,t)
	&\Big)\Big\}\,,\\[3mm]
 F_{S,3}^{{\rm D}}(t)&=&
	2B\Big\{
	0+\frac 1{F^2}\Big(&
	+\Lambda_3^{{\rm D}}\\[2mm]
	&&& \hspace*{-3cm}+\frac \mpis{18} \Bbar(\metas,t)
	&\hspace{-15mm}+\frac 12 (\,\, t+\,\,\mpis  )\Bbar(\mpis,t)
	&
	&-\frac {\mpis}3  \Bbar(\metas,\mpis,t)
	&\Big)\Big\}\,,
\end{array}
\end{equation}
\end{minipage}
where
\begin{equation}\label{eq:results_uds}
\hspace{-0.6mm}\begin{array}{l@{\hspace{1mm}}c@{\hspace{1mm}}l@{\hspace{0mm}}l@{\hspace{0mm}}l@{\hspace{0mm}}l@{\hspace{0mm}}l@{\hspace{0mm}}l@{\hspace{0mm}}l@{\hspace{0mm}}}
 \Lambda_3^{{\rm F}}&=&
	4\Big\{
	\mpis(-6 L_4^r-4 L_5^r&+12 L_6^r &+8 L_8^r)
	&+\mKs (-4 L_4^r+8 L_6^r )
	&+t    (2  L_4^r+ L_5^r)
	&\Big\}\,,\\[3mm]
\Lambda_3^{{\rm C}}&=&
	 4\Big\{
	\mpis(-2 L_4^r-4 L_5^r&+4 L_6^r &+8L_8^r)
	&+\mKs (-4 L_4^r+8 L_6^r )
	&+t     \hspace{12.5mm}L_5^r
	&\Big\}\,,\\[3mm]
\Lambda_3^{{\rm D}}&=&
	 4\Big\{
	\mpis (-4 L_4^r &+ 8 L_6^r)&
	&&+ t\;\,2      L_4^r
	&\Big\}\,.
\end{array}
\end{equation}
The results for the octet and singlet scalar form factors
\begin{equation}\label{eq:octet_singlet}
\begin{array}{lcl}
\langle\pi^i|\bar uu+\bar dd-2\bar ss|\pi^k\rangle&=&\delta^{ik}F_S^8(t)\,,\\[3mm]
\langle\pi^i|\bar uu+\bar dd+\bar ss|\pi^k\rangle &=&\delta^{ik}F_S^0(t)\,,\\[3mm]
\end{array}
\end{equation}
are given in appendix~\ref{app:octet and singlet}.
An interesting observation is the 
absence of low-energy constants in the expression for the disconnected
contribution to the octet form factor due to the vanishing
super-trace of the octet-current.  This allows for its parameter-free 
prediction at this order of the chiral expansion.

\subsection{Numerical results}
The chiral Lagrangian in the 2-flavour theory in~\cite{Gasser:1983yg}
was presented in a different pa\-ra\-mete\-risa\-tion which is related to the one
used here through the relations
$l_3=4(-2 \tilde L_4-\tilde 
L_5+4\tilde L_6 +2\tilde L_8)$ and 
$l_4=4(2\tilde L_4+\tilde L_5)$, hence,
\begin{equation}\label{eq:results}
\begin{array}{l@{\hspace{1mm}}c@{\hspace{1mm}}l@{\hspace{0mm}}l@{\hspace{0mm}}l@{\hspace{0mm}}l@{\hspace{0mm}}l@{\hspace{0mm}}l@{\hspace{0mm}}l@{\hspace{0mm}}l@{\hspace{0mm}}}
\Lambda^{\rm F}_2&=
        &t\;\, l_4^r& &+ 4\mpis\;\, l_3^r\,, \\[3mm]
\Lambda^{\rm C}_2 &= 
        &t ( l_4^r&-8\tilde L_4^r)
	 &+ 4\mpis(l_3^r&+4\tilde L_4^r&-8\tilde L_6^r)\,,\\[3mm]
\Lambda^{\rm D}_2&= 
        &t (      &+8\tilde L_4^r)
	 &+ 4\mpis(       &-4\tilde L_4^r&+8\tilde L_6^r)\,.
\end{array}
\end{equation}
The results for $F_{S,2}^{\rm F}$ and $F_{S,3}^{\rm F}$
agree with the ones found in \cite{Gasser:1983yg,Gasser:1984ux}.
The low-energy parameters 
$\tilde L_4^r$ and $\tilde L_6^r$ 
entering $F_{S,2}^{\rm C}$ and $F_{S,2}^{\rm D}$ are in principle unknown 
from the 2-flavour theory - 
a corollary of the unphysical nature of the connected and the disconnected
contributions considered on their own.  
The corresponding terms 
 are naively  present in the $SU(2)$ chiral Lagrangian
but trace-identities and the use of the equations of motion 
allow to reduce the relevant set of terms, hence  the
above linear combinations defining $l_3$ and $l_4$ in terms of the $\tilde L_i$.

In both the 2- and the 3-flavour theory the disconnected contribution
is suppressed by power counting
relative to the connected contribution and the full form factor, respectively.
Power counting also suggests 
that the contribution of the disconnected contractions to the scalar radius
$\langle r^2\rangle$ defined through
\begin{equation}
F_S(t)=F_S(0)\left(1+\frac 16 \langle r^2\rangle t + O(t^2)\right)\,,
\end{equation}
(and similarly for the octet and singlet charge radius)
can be of about the same magnitude.
Since the tree-level contribution is purely real, the connected
and the disconnected contractions also contribute democratically
to the imaginary part of the form factor 
above the two-pion threshold in the time-like region.

In order to analyse the two contributions to the form factor more
quantitatively we use the lattice estimates for 
the 3-flavour low-energy constants $L_4^r$, $L_5^r$, $L_6^r$ and $L_8^r$
by the RBC/UKQCD lattice collaboration~\cite{Allton:2008pn}\footnote{
See~\cite{Colangelo:2010et} for a summary of determinations of low-energy
parameters in lattice QCD.}
as given in table~\ref{tab:LECs}. Further input parameters are
$F=93$MeV, $m_\pi=0.135$MeV, $m_K=495$MeV and $m_\eta=548$MeV.
With this external input it is straight forward to analyse the 3-flavour case. 
Although $\tilde L_4^r$ and $\tilde L_6^r$ are not known they
can be related to the low-energy constants 
$L_4^r$ and $L_6^r$ by matching the two theories. 
In particular,  we use
\begin{equation}
\begin{array}{lcl}
l_3^r&=&4
	(-8 L_4^r-4L_5^r+16 L_6^r +8L_8^r)-\frac 1{18}\tilde L(\metas)\,,\\[3mm]
l_4^r&=&8L_4^r+4L_5^r-\frac 14 \tilde L(\mKs)\,,\\[3mm]
4\tilde L_4^r-\tilde 8L_6^r&=&4L_4^r-8L_6^r-\frac 54 \frac 1{18}\tilde L(\metas)
			-\frac 1{12} \frac 1N\,,\\[3mm]
\tilde L_4^r&=&L_4^r\,,
\end{array}
\end{equation}
where $\tilde L(m^2)=\frac 1{16\pi^2}\left(1+\ln(m^2/\mu^2)\right)$.
\begin{table}
\begin{center}
\begin{tabular}{lllllll}
\hline\hline&&\\[-4mm]
$L_4^r$&\cite{Allton:2008pn,Colangelo:2010et}&$0.14\times10^{-3}$\\
$L_5^r$&\cite{Allton:2008pn,Colangelo:2010et}&$0.87\times10^{-3}$\\
$L_6^r$&\cite{Allton:2008pn,Colangelo:2010et}&$0.07\times10^{-3}$\\
$L_8^r$&\cite{Allton:2008pn,Colangelo:2010et}&$0.56\times10^{-3}$\\[1mm]
\hline
\hline
\end{tabular}
\caption{Values for the 3-flavour
low energy constants used in the illustration of 
the results. The subtraction scale is $\mu=0.77$GeV.}\label{tab:LECs}
\end{center}
\end{table}
\begin{figure}
\begin{center}
 $N_f=2$\hspace{6cm}$N_f=2+1$\\[3mm]
 \psfrag{Epipisq}[t][t][1][0]{$t\,\rm [GeV^2]$}
 \psfrag{fpipi}[c][t][1][0]{\hspace*{0mm}\mbox{}\normalsize
			${\rm Re}\left(F_{S,2}^{\rm F,C,D}(t)/(2B)\right)$}
 \psfrag{err}[c][t][1][0]{\hspace*{-32mm}\mbox{}\large
			$\Delta^{\rm Re}$}
 \epsfig{scale=0.63,file=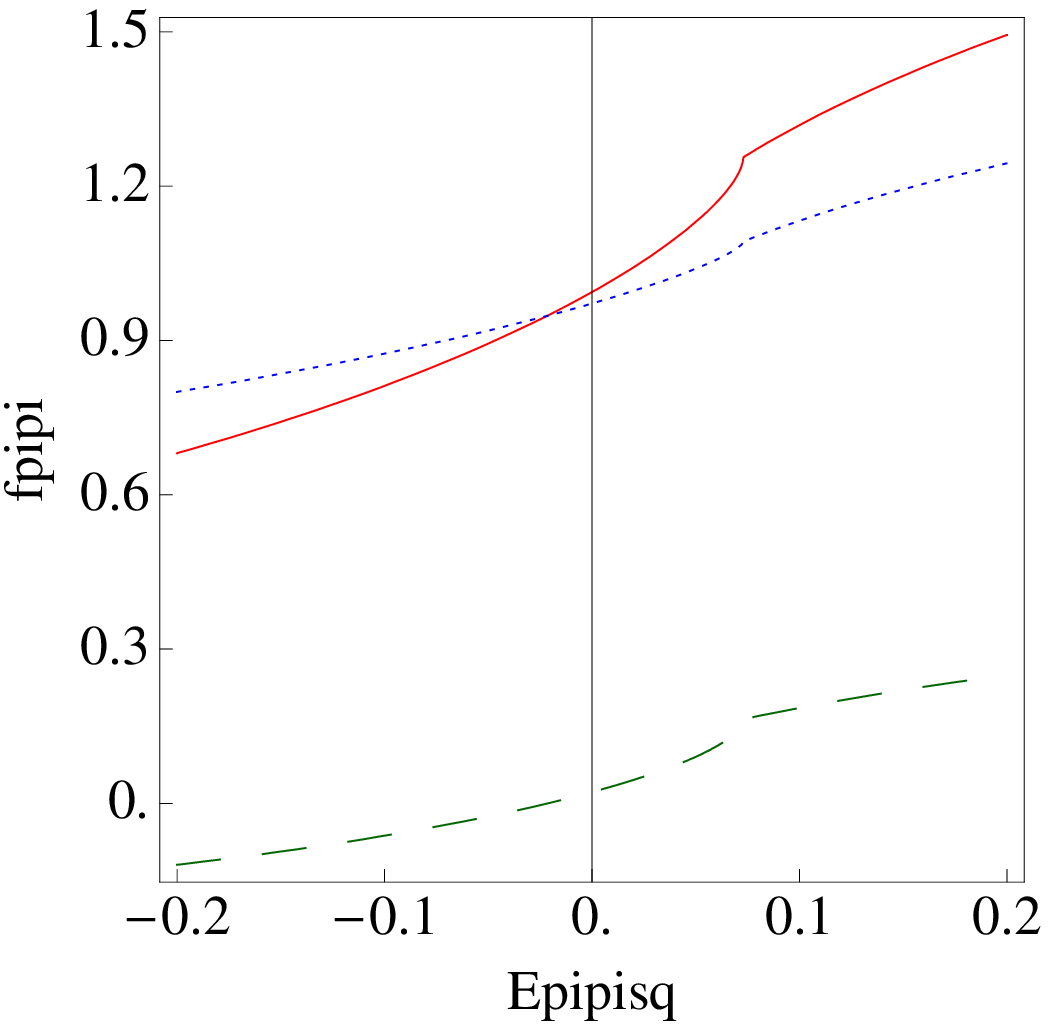}\hspace{8mm}
 \psfrag{fpipi}[c][t][1][0]{\hspace*{0mm}\mbox{}\normalsize
			${\rm Re}\left(F_{S,3}^{\rm F,C,D}(t)/(2B)\right)$}
 \epsfig{scale=0.63,file=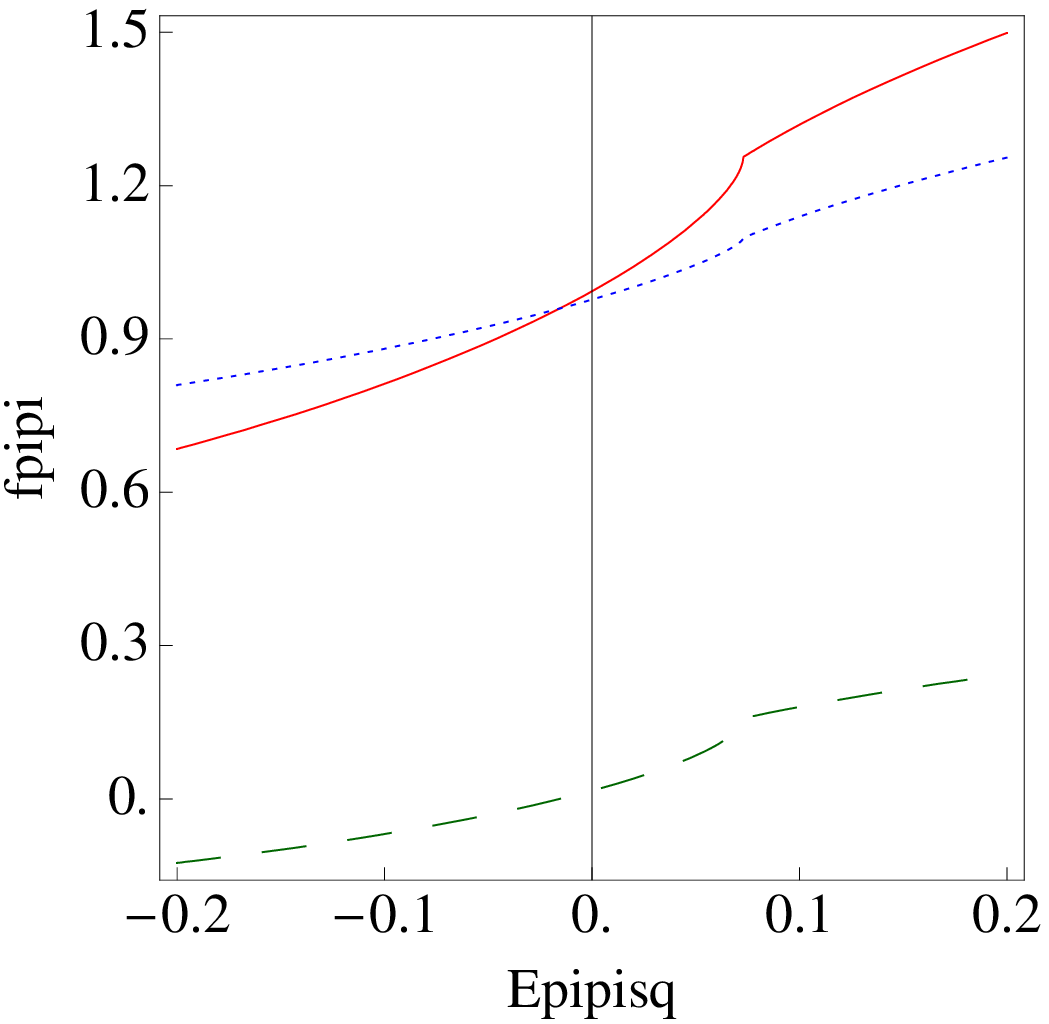}\\[4mm]
 \psfrag{fpipi}[c][t][1][0]{\hspace*{0mm}\mbox{}\normalsize
			${\rm Im}\left(F_{S,2}^{\rm F,C,D}(t)/(2B)\right)$}
 \psfrag{err}[c][t][1][0]{\hspace*{-32mm}\mbox{}\large
			$\Delta^{\rm Im}$}
 \epsfig{scale=0.63,file=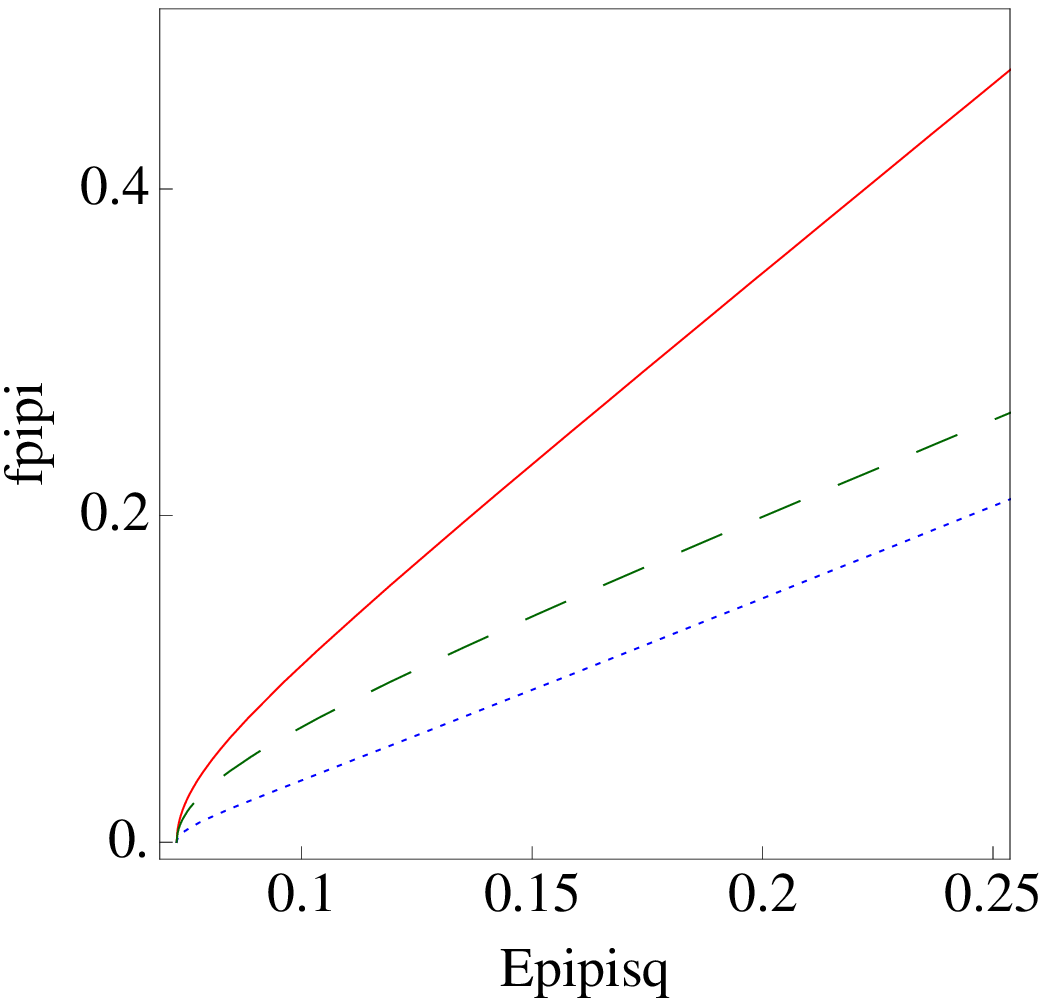}\hspace{8mm}
 \psfrag{fpipi}[c][t][1][0]{\hspace*{0mm}\mbox{}\normalsize
			${\rm Im}\left(F_{S,3}^{\rm F,C,D}(t)/(2B)\right)$}
 \epsfig{scale=0.63,file=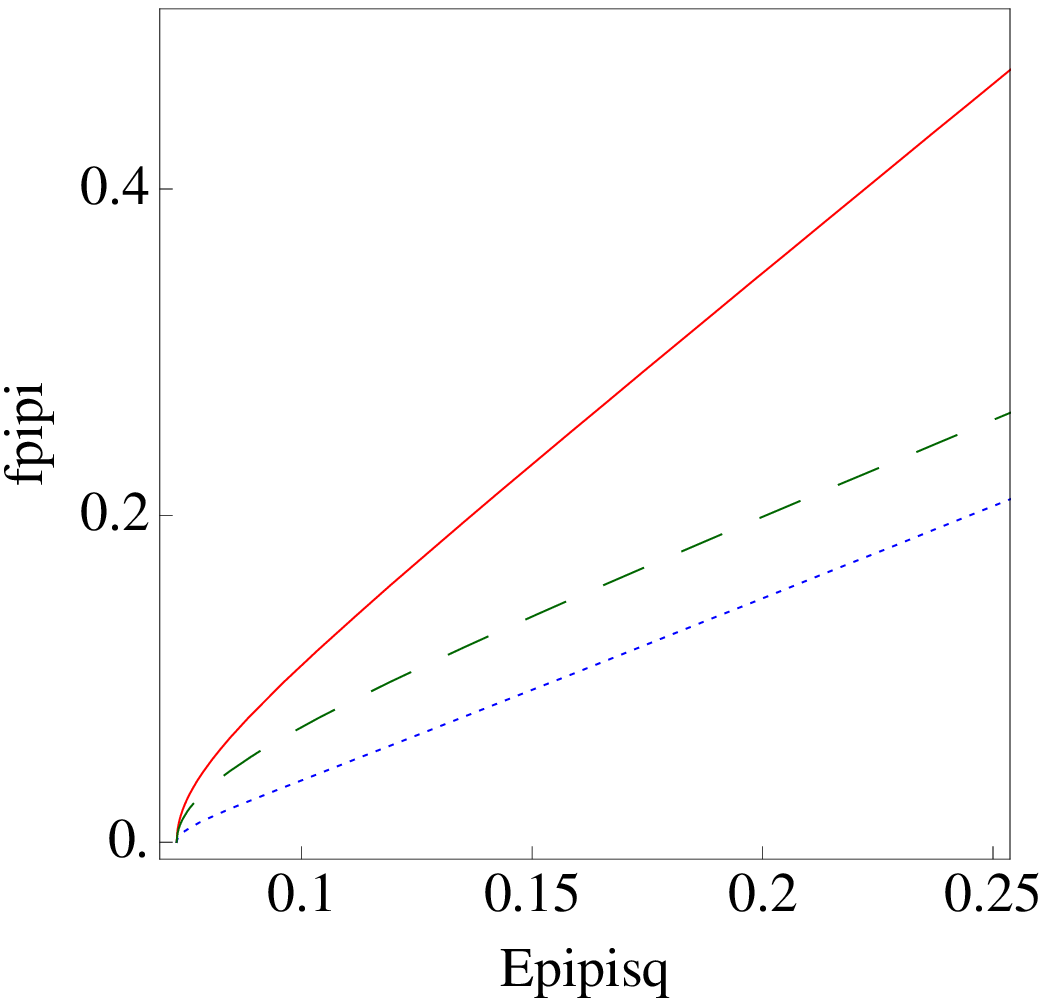}
\end{center}
\caption{The plots show the momentum dependence of the real (top) and 
	imaginary (bottom)
	parts of the scalar form factor for 
	$N_f=2$ (left) and $N_f=2+1$ (right): 
	Full form factor (solid red),
	connected contribution (dotted blue) and disconnected contribution 
	(dashed green).
}\label{fig:results}
\end{figure}
\begin{figure}
\begin{center}
 $N_f=2$\hspace{5.5cm}$N_f=2+1$\\[3mm]
 \psfrag{rsqpi}[c][t][1][0]{\hspace{9mm}$\langle r^2\rangle \;[{\rm fm}^2]$}
 \psfrag{mpisq0}[t][t][1][0]{$m_\pi^2$ [GeV]$^2$}
 \epsfig{scale=0.63,file=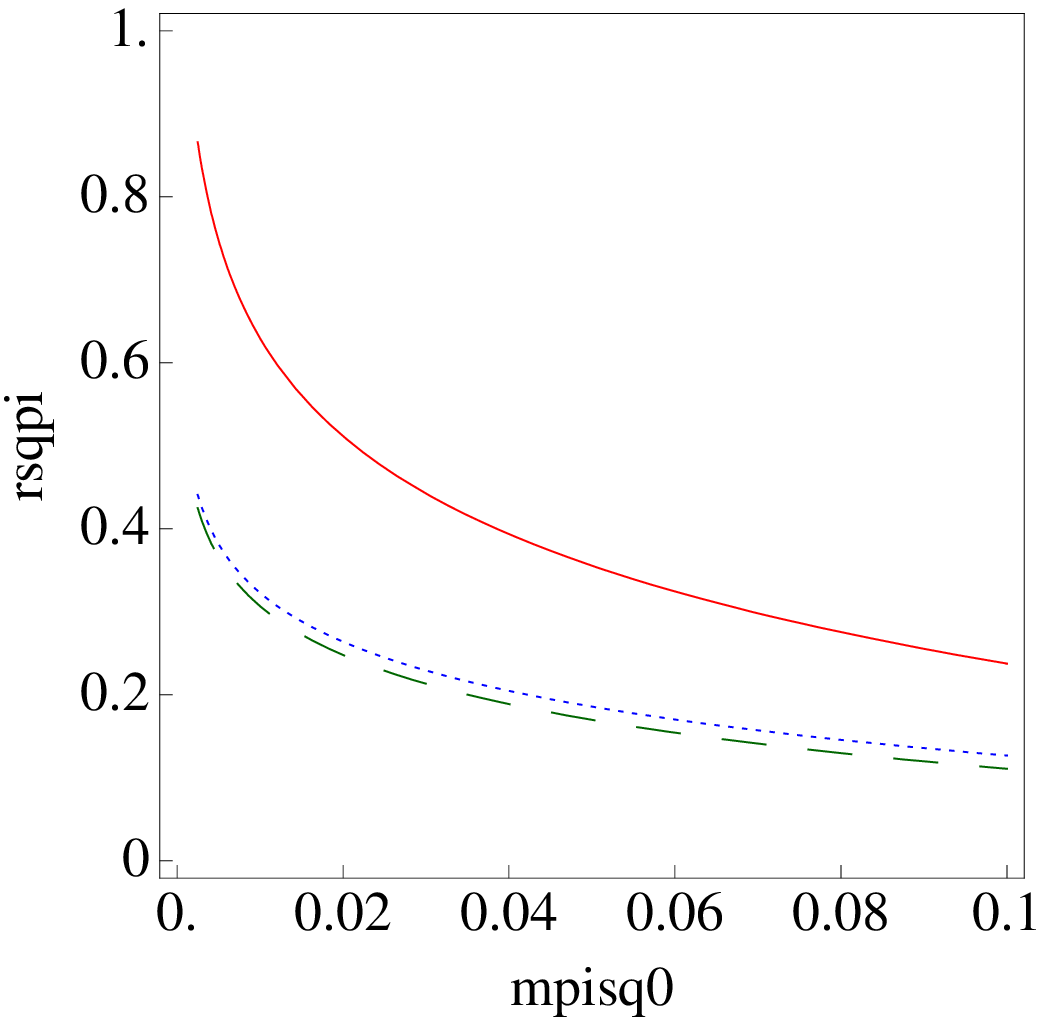}\hspace{4mm}
 \epsfig{scale=0.63,file=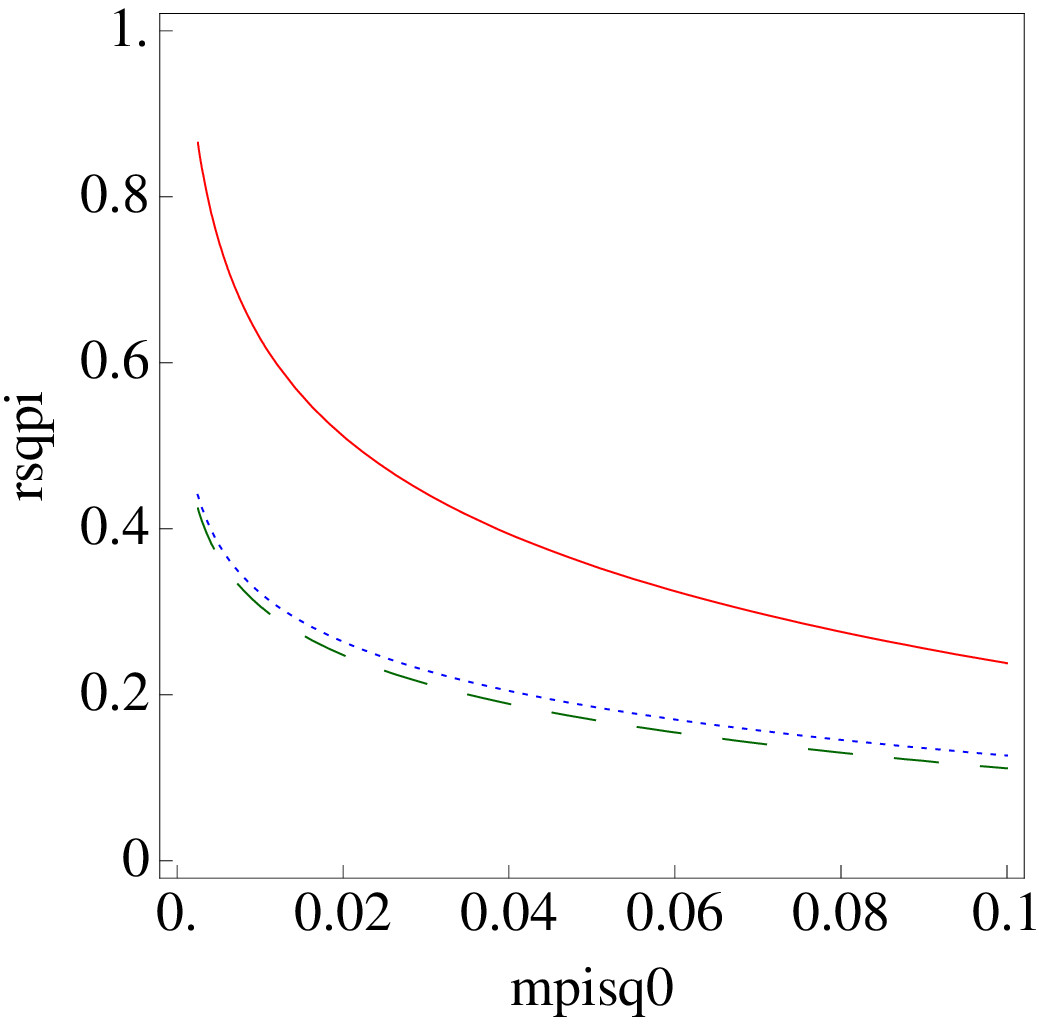}
\end{center}
\caption{The plots show the scalar radius  for 
	$N_f=2$ (left) and $N_f=2+1$ (right) as a function of the
	squared pion mass ($m_K$ is fixed to the physical value
	and $\metas=\frac 43 \mKs-\frac 13\mpis$): 
	full radius (solid red),
	connected contribution (dotted blue) and disconnected contribution 
	(dashed green).
	}
\label{fig:rsq}
\end{figure}
 
The dependence on the momentum of the full, the connected and 
the disconnected form factor are illustrated in the plots in 
figure~\ref{fig:results}. 
As anticipated $F_S^{\rm D}$ (dashed) is suppressed relative 
to $F_S^{\rm C}$ (dotted), predominantly
 due to the absence of a tree-level contribution in the former.
Above threshold the contribution from the 
imaginary part of the form factor sets in and the quark 
disconnected contribution is larger than 50\% over the whole range plotted.
Figure~\ref{fig:rsq} shows the scalar radius 
and as expected the disconnected 
contribution is not suppressed. In fact, both the connected and the
disconnected contractions contribute to roughly equal parts.

The observations made here for the form factor $F_S$ apply qualitatively in the same way 
for the form factors $F_S^{0}$ and $F_S^{8}$.

\section{Applications}
Clearly this study has been motivated by the technical problems encountered
in the computation in lattice QCD of observables receiving contributions from 
quark-disconnected diagrams. Here we highlight three ways in which the
results of this paper might be of use:
\bi
 \item When computing the scalar form factor in lattice QCD with the aim of 
	determining low-energy parameters it is possible to gain valuable 
	information by comparing the 
	dependence of the lattice data for only the
	connected part $F_S^{\rm C}$ on the quark mass and the
	momentum transfer, to the corresponding expressions in the
	effective theory for $N_f=2$ (eq.~(\ref{eq:results_ud}))
	and $N_f=2+1$ (eq.~(\ref{eq:results})).
	A special case in this context is the octet scalar form factor 
	(cf. appendix ~\ref{app:octet and singlet}) since the 
	expression for the disconnected contribution is entirely parameter-free.
 \item If one is interested in the result for the full form factor
	$F_S^{\rm F}$, the only approach which will eventually provide 
	a reliable control of systematic effects is a complete numerical 
	evaluation in lattice QCD. An approximate picture for the form 
	factor can be obtained by combining an evaluation of the connected 
	contribution in lattice QCD with predictions for the disconnected
	contribution by chiral perturbation theory. With the exception of the
	octet form factor the latter of course 
	relies on external input in terms of estimates of the low-energy constants.
 \item In ~\cite{Boyle:2007wg} an argument was given that justifies the 
	use of (partially) twisted boundary conditions for the computation of 
	the pion's iso-spin limit vector form factor. From the flavour structure
	of the correlation functions on the r.h.s. of eq.~(\ref{eq:decomp_quarks})
	it follows that this technique can also be applied to the 
	quark-connected correlator that contributes to the scalar form factor.
\ei
\section{Conclusions and outlook}
This paper provides the expressions for quark-connected and quark-disconnected
Wick contractions, respectively,
 contributing to the pion's scalar form factor and its radius in 
next-to-leading order (partially quenched) chiral perturbation theory. 
While the quark-dis\-con\-nected part in the scalar
form factor is sub-dominant, it turns out to contribute with 
about the same magnitude to the scalar radius as the quark-connected one. 

This result has implications for lattice QCD simulations where 
quark-disconnected
contributions are often neglected \textit{ad hoc}
 because of their typically
bad signal-to-noise ratio. 
Computing quark-disconnected contributions in the effective theory can provide 
power-counting arguments for or against neglecting them for 
QCD-observables where chiral perturbation theory is expected to be a reliable 
effective description of the low-energy properties of the underlying
fundamental theory.
The example of the octet scalar form factor  shows that for a suitable 
choice of the observable a  prediction
of the disconnected contribution in the 
effective theory is possible without loss of predictivity. 

Quark-disconnected contractions contribute to a large class of matrix elements
relevant for the precision phenomenology of the Standard Model. Examples are 
the yet to be fully understood
process $K\to\pi\pi$ or iso-spin breaking effects. 
The technique
employed here and previously in \cite{DellaMorte:2010aq} can potentially
be applied to guide the computation of quark-disconnected contributions in these cases.\\[2mm]

\noindent {\bf Acknowledgments:} The author would like to thank 
Martin L\"uscher for comments
on the manuscript and Michele Della Morte for fruitful discussions.\\[3mm]

\appendix{\noindent{\bf{\large Appendix:}}}
\section{Vertices in the partially quenched theory}
The Feynman rules in partially quenched chiral perturbation theory differ from
the ones in the purely bosonic theory since the commutators of the involved 
fields depend on their \textit{grading} $\eta$. In the following we first provide 
expressions for the traces of products of $SU(N|M)$ generators and then 
give explicit expression for the relevant vertices that derive from 
the Lagrangians~(\ref{eq:L2}) and (\ref{eq:L4}).
\subsection{Traces of $SU(N|M)$ generators}
Assuming the normalisation of the $SU(N|M)$ generators as 
in section~\ref{sec:general_argument} their commutators and anti-commutators are
\begin{equation}
\begin{array}{lclcl}
[T^a,T^b]  &\equiv&T^aT^b-(-)^{\eta_a\eta_b}T^bT^a&=&i C^c_{ab}T^c\\
\{T^a,T^b\}&\equiv&T^aT^b+(-)^{\eta_a\eta_b}T^bT^a&=&\frac 1N g^{ab}+ D^c_{ab}T^c\\
\end{array}
\end{equation}
where $\eta^a$ is the grading of the generator $a$, i.e. 
$\eta^a=0(1)$ if $T^a$ generates a boson(fermion). 
The explicit expressions for the structure constants $C$ and $D$ can be obtained by
projection,
\begin{equation}
\begin{array}{lcl}
C_{ab}^c=-i2\str\Big\{[T^a,T^b]T^k\Big\}g^{kc}\,,\\[4mm]
D_{ab}^c=+\,2\str\Big\{\{T^a,T^b\}T^k\Big\}g^{kc}\,.\\[4mm]
\end{array}
\end{equation}
With these definitions the following results for the super-traces of 
products of $SU(N|M)$-generators are obtained:
\begin{equation}
\begin{array}{lcl}
 \str\left\{T^a T^b\right\} &=& \frac 12 g^{ab}\,,\\[4mm]
 \str\left\{T^a T^b T^c\right\} &=&\frac 14
		\left(
			iC^o_{ab}+D^o_{ab}
		\right)g^{oc}\,, \\[4mm]
 \str\left\{T^a T^b T^cT^d\right\} &=&\frac 14
 \str\left\{
    \left(\frac 1N g^{ab}+ \left(iC^k_{ab}+D^k_{ab}\right)T^k\right)
	\right.\\[4mm]
&&\hspace{06.0mm}\left.    
	\times\left(
	  \frac 1N g^{cd}+ \left(iC^l_{cd}+D^l_{cd}\right)T^l
	\right)
	\right\}\\[6mm]
	&=&\frac 14 \frac 1N g^{ab}g^{cd}+
	\frac 18\left(i C_{ab}^k+D_{ab}^k\right)g^{kl}
		\left(i C_{cd}^l+D_{cd}^l\right)\,,\\[4mm]
 \str\left\{T^a T^b T^cT^dT^e\right\} &=&\frac 14
 \str\left\{
    \left(\frac 1N g^{ab}+ \left(iC^k_{ab}+D^k_{ab}\right)T^k\right)
	\right.\\[4mm]
 &&\hspace{6.3mm}\left.   
	\times\left(
	\frac 1N g^{cd}+\left(iC^l_{cd}+D^l_{cd}\right)T^l
	\right)T^e
	\right\}\\[6mm]
	&=&\frac 1{8N}g^{ab}\left(iC^l_{cd}+D^l_{cd}\right)g^{le}+
	   \frac 1{8N}g^{cd}\left(iC^k_{ab}+D^k_{ab}\right)g^{ke}\\[4mm]
	&&  \;\;\;+ \frac 1{16}
	 	\left(iC^k_{ab}+D^k_{ab}\right)
		\left(iC^l_{cd}+D^l_{cd}\right)
		\left(iC^o_{kl}+D^o_{kl}\right)g^{oe}\,.
\end{array}
\end{equation}
\subsection{Vertices}\label{app:vertices}
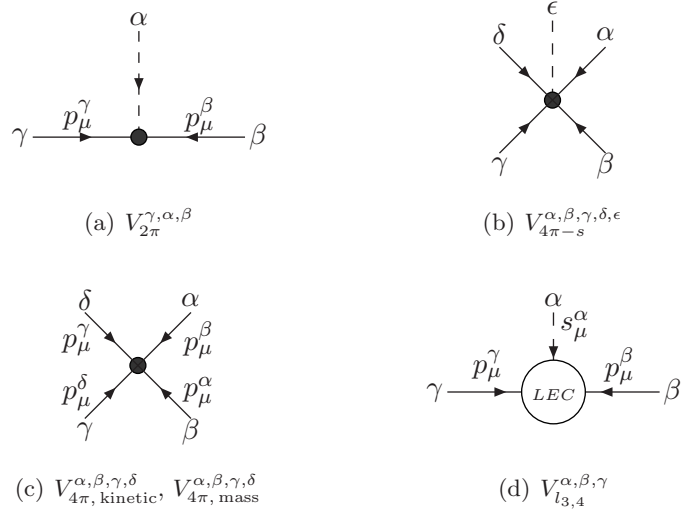
\begin{figure}
 \begin{center}
 \subfigure[$V_{2\pi}^{\gamma,\alpha,\beta}$]{
    \begin{picture}(80,40)(-40,-21)
    \Text(0,45)[c]{$\alpha$}
    \Text(-45,0)[c]{$\gamma$}
    \Text(+45,0)[c]{$\beta$}
    \Text(+23,+08)[c]{$p_\mu^\beta$}
    \Text(-23,+08)[c]{$p_\mu^\gamma$}
    \DashArrowLine(0,40)(0,0){5}
    \ArrowLine(-40,0)(0,0)
    \ArrowLine(+40,0)(0,0)
    \GCirc(0,0){3}{.2}
    \end{picture}
    \label{fig:feyn_a}
  }\hspace{23mm}
  \subfigure[$V_{4\pi-s}^{\alpha,\beta,\gamma,\delta,\epsilon}$]{
    \begin{picture}(80,40)(-40,-25)
    \DashLine(0,10)(0,40){5}
    \GCirc(0,10){3}{.2}
    \Text(0,45)[c]{$\epsilon$}
    \Text(+20,+35)[c]{$\alpha$}
    \Text(+20,-15)[c]{$\beta$}
    \Text(-20,-15)[c]{$\gamma$}
    \Text(-20,+35)[c]{$\delta$}
    \ArrowLine(-20,+30)(+0,+10)
    \ArrowLine(-20,-10)(+0,+10)
    \ArrowLine(+20,+30)(+0,+10)
    \ArrowLine(+20,-10)(+0,+10)
    \end{picture}
    \label{fig:feyn_b}
   }\\[6mm]
  \subfigure[$V_{4\pi,\,{\rm kinetic}}^{\alpha,\beta,\gamma,\delta}$,~$V_{4\pi,\,{\rm mass}}^{\alpha,\beta,\gamma,\delta}$]{
    \begin{picture}(100,60)(-50,-25)
    \GCirc(0,10){3}{.2}
    \Text(+20,+35)[c]{$\alpha$}
    \Text(+20,-15)[c]{$\beta$}
    \Text(-20,-15)[c]{$\gamma$}
    \Text(-20,+35)[c]{$\delta$}
    \ArrowLine(-20,+30)(+0,+10)
    \ArrowLine(-20,-10)(+0,+10)
    \ArrowLine(+20,+30)(+0,+10)
    \ArrowLine(+20,-10)(+0,+10)
    \Text(+23,+00)[c]{$p_\mu^\alpha$}
    \Text(-23,+00)[c]{$p_\mu^\delta$}
    \Text(+23,+20)[c]{$p_\mu^\beta$}
    \Text(-23,+20)[c]{$p_\mu^\gamma$}
    \end{picture}
    \label{fig:feyn_c}
   } \hspace{15mm}
  \subfigure[$V_{l_{3,4}}^{\alpha,\beta,\gamma}$]{
    \begin{picture}(100,60)(-50,-25)
    \Text(+0,+35)[c]{$\alpha$}
    \Text(+45,-00)[c]{$\beta$}
    \Text(-45,-00)[c]{$\gamma$}
    \Text(+8,+25)[c]{$s^\alpha_\mu$}
    \Text(+25,+10)[c]{$p^\beta_\mu$}
    \Text(-25,+10)[c]{$p^\gamma_\mu$}
    \DashArrowLine(0,30)(0,0){5}
    \ArrowLine(-40,+0)(+0,+0)
    \ArrowLine(+40,+0)(+0,+0)
    \GCirc(0,0){12}{1}
    \Text(+0,-1)[c]{\tiny$ LEC$}
    \end{picture}
    \label{fig:feyn_d}
   }
  \end{center}
\caption{Vertices entering the computation of the scalar form factor.}
\end{figure}
\subsubsection{$2\pi-s$-vertex}
The vertex illustrated in figure~\ref{fig:feyn_a} is
\begin{equation}
\begin{array}{r@{\hspace{1mm}}c@{\hspace{1mm}}l}
V_{2\pi s}^{\gamma,\alpha,\beta}&=&- i2 B 
	\Big\{
	(-1)^{\eta^\alpha(\eta^\beta+\eta^\gamma)}
	\Big(     \Str\left\{T^\alpha T^\gamma T^\beta\right\}+
	(-1)^{\eta^\beta \eta^\gamma}\Str\left\{T^\alpha T^\beta T^\gamma\right\}	
	\\[3mm]
	&&\hspace*{34mm}+
	     \Str\left\{T^\gamma T^\beta T^\alpha\right\}+
	(-1)^{\eta^\beta \eta^\gamma}\Str\left\{T^\beta T^\gamma T^\alpha\right\}
	\Big)
	\Big\}\,
\end{array}
\end{equation}
\subsubsection{$4\pi$  and $4\pi-s$ vertices}
The $4\pi-s$ vertex in figure~\ref{fig:feyn_b} is
\begin{equation}\label{eq:4pis}
 \begin{array}{rcl}
 V_{4\pi s}^{\gamma,\delta,\epsilon,\alpha,\beta}&=&
	 i\frac 23  \frac {B}{F^2} 
          \sum\limits_{\mathcal{P}}
	\mathcal{S}(\sigma(\alpha),\sigma(\beta),\sigma(\gamma),\sigma(\delta))\\
&&\qquad\times	\Big\{
 		(-1)^{\eta^\epsilon(\eta^\alpha+\eta^\beta+\eta^\gamma+\eta^\delta)}
		\Str\left\{T^{\epsilon}T^{\sigma(\alpha)} T^{\sigma(\beta)}
	                   T^{\sigma(\gamma)} T^{\sigma(\delta)}\right\}\\[3mm]
&&\qquad\hspace{3.55cm}		+\Str\left\{T^{\sigma(\alpha)} T^{\sigma(\beta)}
	                   T^{\sigma(\gamma)} T^{\sigma(\delta)}T^{\epsilon}\right\}
	\Big\}
		\,,
 \end{array}
\end{equation}
and the vertex deriving from the mass term as in figure ~\ref{fig:feyn_c} is
\begin{equation}\label{eq:4pi}
 \begin{array}{rcl}
 V_{4\pi\,M}^{\gamma,\delta,\alpha,\beta}&=&
	 i\frac {2}{3}  \frac{B}{F^2} 
          \sum\limits_{\mathcal{P}}
	\mathcal{S}(\sigma(\alpha),\sigma(\beta),\sigma(\gamma),\sigma(\delta))
		\Str\left\{MT^{\sigma(\alpha)} T^{\sigma(\beta)}
	                   T^{\sigma(\gamma)} T^{\sigma(\delta)}\right\}\,.
 \end{array}
\end{equation}
Here $\mathcal{P}$ indicates that the sum is over all permutations $\sigma$
of the 
indices. 
Starting from $(\alpha,\beta,\gamma,\delta)$ a given permutation
$(\sigma(\alpha),\sigma(\beta),\sigma(\gamma),\sigma(\delta))$ may be 
reached by a series of exchanges of pairs of neighbouring indices. 
$\mathcal{S}(\alpha,\beta,\gamma,\delta)$ is the product of signs
that are picked up in each such exchange
of two indices depending on its grading. For example,
for $(\sigma(\alpha),\sigma(\beta),\sigma(\gamma),\sigma(\delta))=
	(\gamma,\beta,\alpha,\delta)$, 
$S(\gamma,\beta,\alpha,\delta)=
(-1)^{\eta^\gamma(\eta^\beta+\eta^\alpha)+\eta^\beta\eta^{\alpha}}$.

The four point vertex deriving from the kinetic term is
\begin{equation}
 \begin{array}{rcl}
 V_{4\pi,\,{\rm kin}}^{\gamma,\delta,\alpha,\beta}&=&
	 i\frac {1}{F^2} 
          \sum\limits_{\mathcal{P}}
	\mathcal{S}(\sigma(\alpha),\sigma(\beta),\sigma(\gamma),\sigma(\delta))
		P(\sigma(\alpha),\sigma(\beta),\sigma(\gamma),\sigma(\delta))\\[3mm]
&& \qquad\qquad\qquad\times	\Str\left\{T^{\sigma(\alpha)} T^{\sigma(\beta)}
	                   T^{\sigma(\gamma)} T^{\sigma(\delta)}\right\}
		\,,
 \end{array}
\end{equation}
where 
$P(a,b,c,d)=	  \frac 23 p^a\cdot p^b 
		- \frac 13 p^a\cdot p^c
		+ \frac 13 p^a\cdot p^d
		- \frac 13 p^b\cdot p^d
		-          p^b\cdot p^c
		+ \frac 23 p^c\cdot p^d$.
In the case where all external legs are bosons the known vertices of $SU(N)$
chiral perturbation theory are recovered.
\subsubsection{Counter terms}
The counter terms needed derive from $\mathcal{L}^{(4)}$ as given in 
eq.~(\ref{eq:L4}) and the corresponding vertices with the indices associated to the external
legs as illustrated in figure~\ref{fig:feyn_d} are 
(we assume here that the external legs fulfil $\eta^\beta=\eta^\gamma=0$),

\begin{equation}
 \begin{array}{rcl}
	V_{L_4}^{\gamma,\alpha,\beta}&=&
	-i\frac {32 L_4 B}{F^2}\,p\cdot p^{\,\prime} \,\Str \left\{T^\alpha\right\}
			       \Str \left\{T^\beta T^\gamma+
				T^\gamma T^\beta\right\}\,,\\[3mm]
	V_{L_5}^{\gamma,\alpha,\beta}&=&
	-i\frac {32 L_5 B}{F^2} \,p\cdot p^{\,\prime}\,\Str \left\{T^\alpha\left(
			       T^\beta T^\gamma+ T^\gamma T^\beta\right)\right\}\,,\\[3mm]
	V_{L_6}^{\gamma,\alpha,\beta}&=&
	+i\frac {128 L_6 B}{F^2} \Big\{
	\Str \left\{M\right\}\hspace{1.0mm}
	\Str\left\{T^\alpha\left(T^\beta T^\gamma+T^\gamma T^\beta\right)\right\}\\[2mm]
       &&\qquad\hspace*{7.5mm}+\Str \left\{T^\alpha\right\}
        \Str\left\{M\left(T^\beta T^\gamma+T^\gamma T^\beta\right)\right\}\Big\}\,,\\[3mm]
	V_{L_8}^{\gamma,\alpha,\beta}&=&
	-i\frac {64 L_8 B}{F^2} 
	\Big\{
	\Str\left\{T^\alpha M T^\beta T^\gamma\right\}
	 +\Str\left\{T^\alpha M T^\gamma T^\beta\right\}\\[3mm]
	&&\qquad\hspace*{4mm} +\left(\Str\left\{MT^\alpha T^\beta T^\gamma\right\}
	+ \Str\left\{MT^\alpha T^\gamma T^\beta \right\}\right)\\[2mm]
	&&\qquad\hspace*{2mm}+2\left(\Str\left\{MT^\beta T^\alpha T^\gamma\right\}
	+ \Str\left\{MT^\gamma T^\alpha T^\beta \right\}\right)
	\Big\}\,.
 \end{array}
\end{equation}
\section{Results for the octet and singlet form factors}\label{app:octet and singlet}
Here we provide the results for the octet and the singlet 
scalar form factor as defined in eq.~(\ref{eq:octet_singlet}),
\begin{equation}\label{eq:results_udm2s}
\hspace{-0.00cm}\begin{array}{l@{\hspace{0mm}}c@{\hspace{1mm}}l@{\hspace{0mm}}l@{\hspace{0mm}}l@{\hspace{0mm}}l@{\hspace{0mm}}l@{\hspace{0mm}}l@{\hspace{0mm}}l@{\hspace{0mm}}}
 F_{S,3}^{{\rm F},8}(t)&=&
	2B\Big\{
	1+\frac{1}{F^2}\Big(\mathcal{A}
	&+\Lambda_3^{{\rm F}\,,8}\\[2mm]
	&&&&\hspace*{-4cm} -\frac{\mpis}{6}\Bbar(\metas,t)
	&+\frac 12 ( 2t-\;\mpis)\Bbar(\mpis,t)
	&-\frac t4 \Bbar(\mKs,t)
	&
	&
	\Big)\Big\}\,,\\[3mm]
 F_{S,3}^{{\rm C},8}(t)&=&
	2B\Big\{
	1+\frac{1}{F^2}\Big(\mathcal{A}
	&+\Lambda_3^{{\rm C}\,,8}\\[2mm]
	&&&&\hspace*{-4cm} & +\frac 12 ( \;t -2\mpis )\Bbar(\mpis,t)
	&+\frac t4 \Bbar(\mKs ,t)
	&+\frac {\mpis}3 \Bbar(\metas,\mpis,t)
	&\Big\}\,,\\[3mm]
 F_{S,3}^{{\rm D},8}(t)&=&
	2B\Big\{
	\hspace{7mm}\frac{1}{F^2}\Big(&+\Lambda_3^{{\rm D}\,,8}\\[2mm]
	&&&&\hspace*{-4cm}  -\frac \mpis{6} \Bbar(\metas,t)
	&+\frac 12 (\;t + \;\mpis)\Bbar(\mpis,t)
	&-\frac t2 \Bbar(\mKs ,t)
	&-\frac {\mpis}3  \Bbar(\metas,\mpis,t)
	&\Big\}\,,
\end{array}
\end{equation}
and
\begin{equation}\label{eq:results_uds}
\hspace{-0.cm}\begin{array}{l@{\hspace{1mm}}c@{\hspace{1mm}}l@{\hspace{0mm}}l@{\hspace{0mm}}l@{\hspace{0mm}}l@{\hspace{0mm}}l@{\hspace{0mm}}l@{\hspace{0mm}}l@{\hspace{0mm}}}
 F_{S,3}^{{\rm F},0}(t)&=&
	2B\Big\{
	1+\frac{1}{F^2}\Big(\mathcal{A}
	&+\Lambda_3^{{\rm F}\,,0}\\[2mm]
	&&&&\hspace{-4cm} +\frac{\mpis}{6}\Bbar(\metas,t)
	&+ \frac 12(2t-\;{\mpis})\Bbar(\mpis,t)
	&+\frac t2 \Bbar(\mKs,t)
	&
	&
	\Big\}\,,\\[3mm]
F_{S,3}^{{\rm C},0}(t)&=&
	2B\Big\{
	1+\frac{1}{F^2}\Big(\mathcal{A}
	&+\Lambda_3^{{\rm C}\,,0}\\[2mm]
	&&&&\hspace{-4cm}& + \frac 12(\;t-2\mpis)\Bbar(\mpis,t)
	&+\frac t4 \Bbar(\mKs ,t)
	&+\frac {\mpis}3 \Bbar(\metas,\mpis,t)
	&\Big\}\,,\\[3mm]
F_{S,3}^{{\rm D},0}(t)&=&
	2B\Big\{
	\hspace{7mm}\frac{1}{F^2}\Big(&+\Lambda_3^{{\rm D}\,,1}\\[2mm]
	&&&&\hspace{-4cm} +\frac \mpis{6} \Bbar(\metas,t)
	&+\frac 12 (\;t+ \;\mpis)\Bbar(\mpis,t)
	&+\frac t4 \Bbar(\mKs ,t)
	&-\frac {\mpis}3  \Bbar(\metas,\mpis,t)
	&\Big\}\,.
\end{array}
\end{equation}
These form factors have the following dependence on low-energy constants:
\begin{equation}\label{eq:results_uds}
\begin{array}{l@{\hspace{1mm}}c@{\hspace{1mm}}l@{\hspace{0mm}}l@{\hspace{0mm}}l@{\hspace{0mm}}l@{\hspace{0mm}}l@{\hspace{0mm}}l@{\hspace{0mm}}l@{\hspace{0mm}}}
 \Lambda_3^{{\rm F},8}&=&
	4\Big\{
	\mpis(-2 L_4^r-4 L_5^r+4 L_6^r +8L_8^r)+
	\mKs (-4 L_4^r+8 L_6^r )+
	t    L_5^r
	&\Big\},\\[3mm]
\Lambda_3^{{\rm C}\,,8}&=&
	4\Big\{
	\mpis(-2 L_4^r-4 L_5^r+4 L_6^r +8L_8^r)+
	\mKs (-4 L_4^r+8 L_6^r )+
	t     L_5^r
	&\Big\},\\[3mm]
\Lambda_3^{{\rm D}\,,8}&=&0\,,
\end{array}
\end{equation}
and
\begin{equation}\label{eq:results_uds}
\hspace{-0.00cm}\begin{array}{l@{\hspace{1mm}}c@{\hspace{1mm}}l@{\hspace{0mm}}l@{\hspace{0mm}}l@{\hspace{0mm}}l@{\hspace{0mm}}l@{\hspace{0mm}}l@{\hspace{0mm}}l@{\hspace{0mm}}}
 \Lambda_3^{{\rm F},0}&=&
	4\Big\{
	\mpis(-8 L_4^r-4 L_5^r&+16 L_6^r +8L_8^r)+
	&\mKs (-4 L_4^r+8 L_6^r )
	&+t    (3  L_4^r+ L_5^r)
	&\Big\}\,,\\[3mm]
\Lambda_3^{{\rm C},0}&=&
	4\Big\{
	\mpis(-2 L_4^r-4 L_5^r&+\,4 L_6^r +8L_8^r)+
	&\mKs (-4 L_4^r+8 L_6^r )
	&+t  \hspace{12mm}   L_5^r
	&\Big\}\,,\\[3mm]
\Lambda_3^{{\rm D},0}&=&
	4\Big\{
	\mpis (-6 L_4^r &+ 12 L_6^r)
	&&+t\,3       L_4^r
	&\Big\}\,.
\end{array}
\end{equation}
Note that $F_S^{\rm D,\,8}$ is an entirely parameter-independent prediction for the
disconnected contribution. 
\section{Kinematical functions}\label{app:kinfunctions}
Here we provide the expressions for the kinematical functions
$\Abar(m^2)$ (tadpole),  $\bar B(m^2,\tilde m^2,t)$ and $\Bbar(m^2,t)$ (both
unitary contributions)
\begin{equation}
\begin{array}{rcl}
\Abar(m^2)&=&-\frac {m^2}{16\pi^2}\ln(m^2/\mu^2)\,,\\[3mm]
\Bbar(m^2,\tilde m^2,t)&=&
	-\frac 1{16\pi^2} \frac{m^2\ln(m^2/\mu^2)-\tilde m^2\ln (\tilde m^2/\mu^2)}
				{m^2-\tilde m^2}\\
	&&+\frac 1{32\pi^2} \left(2+\left(-\frac{\Delta}{t}+\frac{\Sigma}{\Delta}\right)
		\ln(m^2/\tilde m^2)
		-\frac \nu t \ln\frac{(t+\nu)^2-\Delta^2}{(t-\nu)^2-\Delta^2}\right)\,,
		\\[3mm]
\Bbar(m^2,t)&=&\lim\limits_{\tilde m^2\to m^2}\Bbar(m^2,\tilde m^2,t)\,,\\
\end{array}
\end{equation}
where, as in e.g. \cite{Gasser:1984gg}, 
$\Delta=m^2-\tilde m^2$, $\Sigma=m^2+\tilde m^2$ and 
$\nu^2=(t-(m+\tilde m)^2)(t-(m-\tilde m)^2))$.

\bibliographystyle{JHEP}
\bibliography{disco2}

\end{document}